\documentclass[12pt]{article}
\usepackage{graphicx}

\newcommand\pubdate{\today}

\def\Title#1{\begin{center} {\Large #1 } \end{center}}
\def\Author#1{\begin{center}{ \sc #1} \end{center}}
\def\Address#1{\begin{center}{ \it #1} \end{center}}

\newcommand\pubblock{\rightline{\begin{tabular}{l}  \\ 
         \pubdate  \end{tabular}}}
\newenvironment{Abstract}{\begin{quotation}  }{\end{quotation}}
\newenvironment{Presented}{\begin{quotation} \begin{center} 
             PRESENTED AT\end{center}\bigskip 
      \begin{center}\begin{large}}{\end{large}\end{center} \end{quotation}}

\begin{document}
\begin{titlepage}
 \pubblock
\vfill
\Title{Jet substructure measurements in CMS}
\vfill
\Author{Christophe Royon}
\Address{Department of Physics and Astronomy, The University of Kansas, Lawrence, USA}
\vfill
\begin{Abstract}
Various recent measurements from the CMS collaboration related to the study of hadronic jets substructure in proton collisions at 13 TeV with the CMS experiment are presented, namely the generalized angular studies in dijet and $Z+$jet events and the measurement of the primary Lund jet plane density.
\end{Abstract}
\vfill
\begin{Presented}
DIS2023: XXX International Workshop on Deep-Inelastic Scattering and
Related Subjects, \\
Michigan State University, USA, 27-31 March 2023 \\
     \includegraphics[width=9cm]{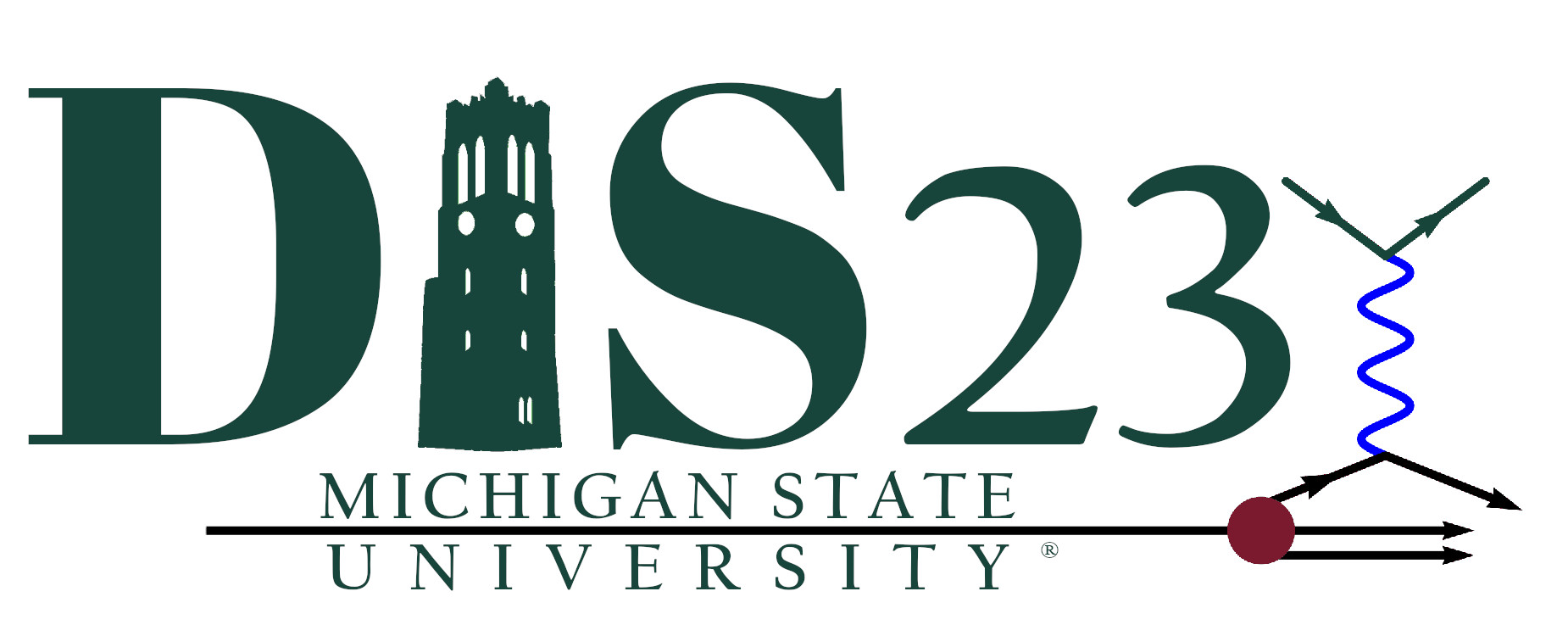}
\end{Presented}
\vfill
\end{titlepage}

Measuring jet substructures is fundamental to compare them with QCD predictions. The idea is to map the jet constituents onto physically meaningful observables. 
We can distinguish between fragmentation functions (the leading hadrons are identified), the classic jet shapes (such as the thrust), and groomed variables (where we want to remove the effects of soft gluon emissions for instance during hadronization).
In this short report, we will mention two different measurements from the CMS collaboration, namely the generalized angular studies in dijet and $Z+$jet events and the measurement of the primary Lund jet plane density.

\section{Generalized angular properties in $Z+$jet and dijet events}

In order to study jet substructures in  $Z+$jet and dijet events, we define new observables~\cite{observables}
$\lambda_{\beta}^{\kappa} = \Sigma_i z_i^{\kappa} \left( \frac{\Delta R_i}{R} \right)^{\beta} $ 
and $z_i = \frac{p_{Ti}}{\Sigma_j P_{Tj}} $
where the sum stands over all jet constituents, $z_i$ is the  jet fractional transverse momentum carried by  the jet component $i$,
$\Delta R_i=\sqrt{(\Delta y_i)^2+(\Delta \phi_i)^2}$ between the jet axis and the jet constituent.
$\beta$ and $\kappa$ are some parameters controlling the momentum and angular distributions. High $\beta$ values enhance angular effects whereas high $\kappa$ values momentum effects. The CMS collaboration studied Les Houches Angularity ($\lambda_{0.5}^1$), the width ($\lambda_{1}^1$), the thrust ($\lambda_{2}^1$),  the multiplicity ($\lambda_{0}^0$) and  $(p_T^D)^2$ ($\lambda_{0}^2$)~\cite{cms1}. Two different samples are used to distinguish between quark and gluon induced jets. $Z+$jet is a quark enhanced sample while dijets are gluon enhanced, especially for central dijets.

\begin{figure}
\centerline{%
\includegraphics[width=9.cm]{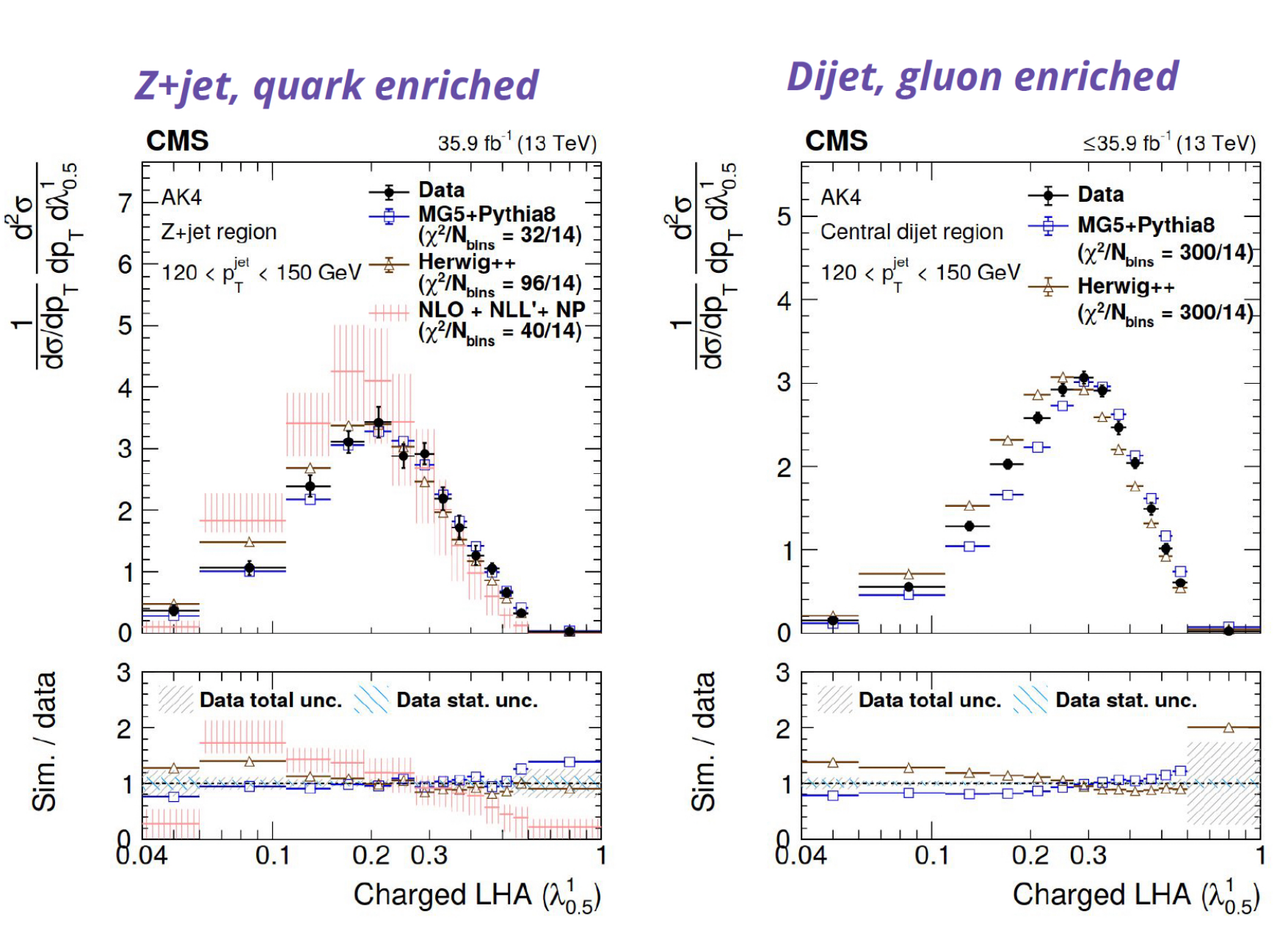}}
\caption{Les Houches angularity observable $\kappa=1$, $\lambda=0.5$ for $Z+$jet (left) and dijet events (right).}
\label{fig1}
\end{figure}

As an example, the Les Houches angularity observable ($\kappa=1$, $\lambda=0.5$) is shown in Fig.~\ref{fig1}.
Data are unfolded to particle level.
MG5$+$PYTHIA~\cite{pythia8} and HERWIG$++$~\cite{herwig7} describe quark-enriched data well, and envelop the gluon-enriched data.
For Z$+$jet events, the resummation at NLL matched to fixed-order NLO matrix elements, with non-perturbative corrections from
SHERPA~\cite{sherpa} is not in perfect agreement with data.

\begin{figure}
\centerline{%
\includegraphics[width=4.5cm]{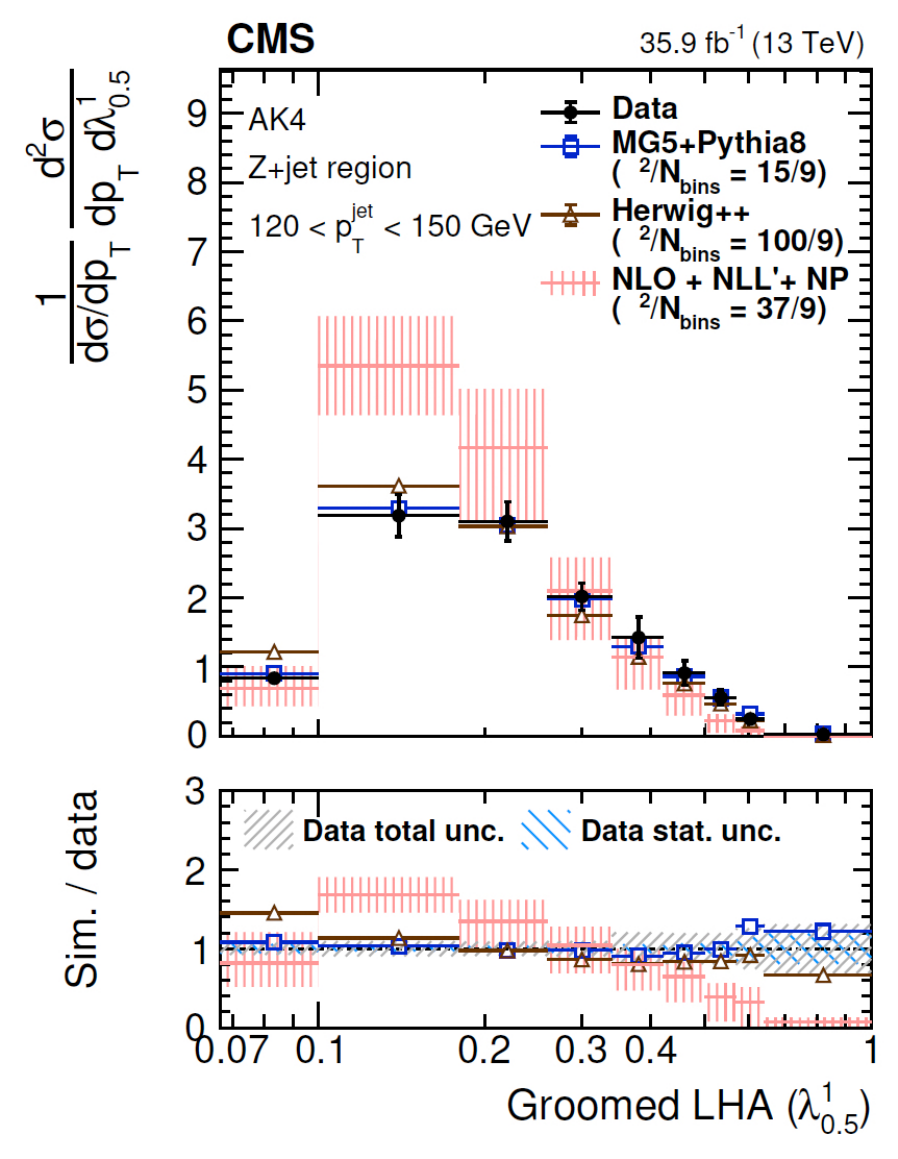}
\includegraphics[width=4.5cm]{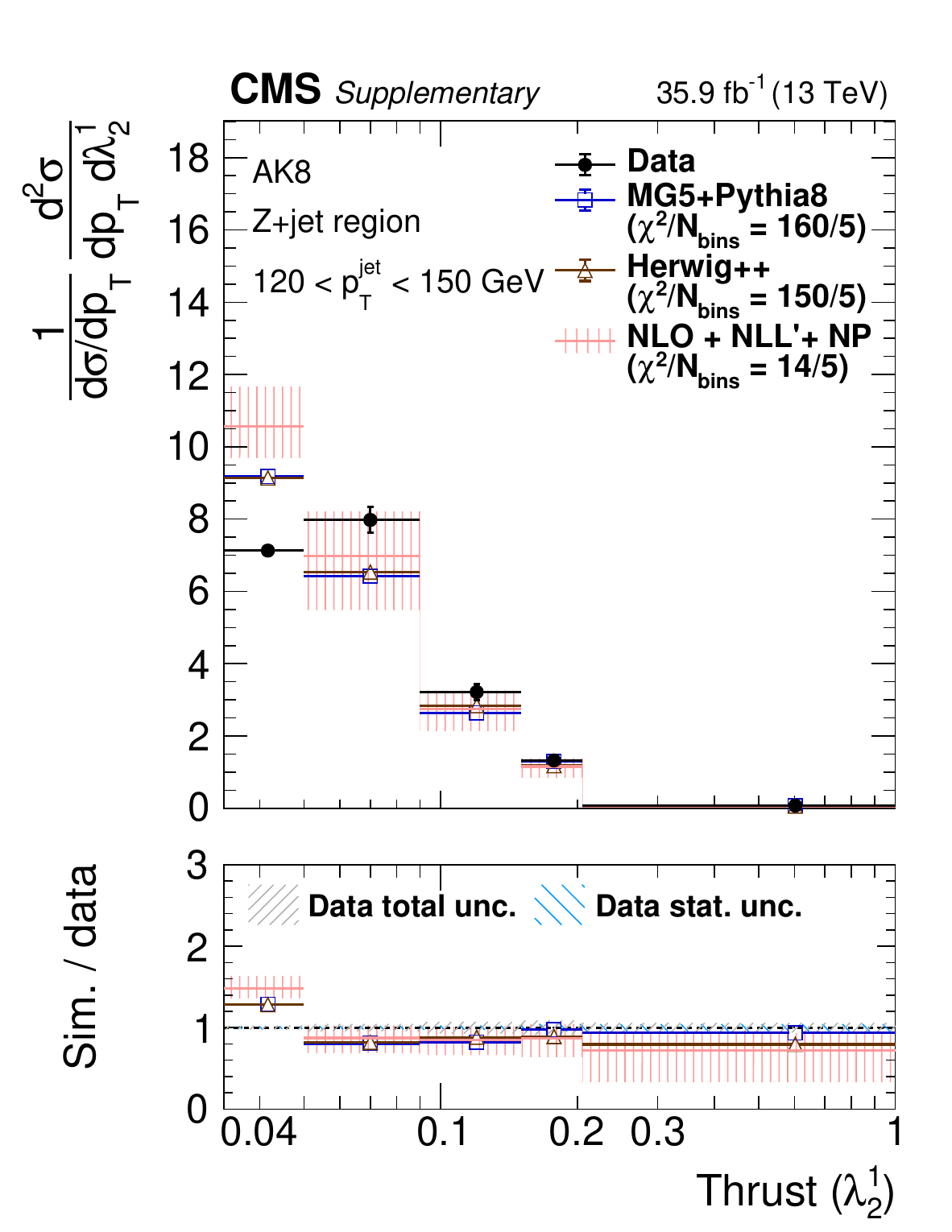}
\includegraphics[width=4.5cm]{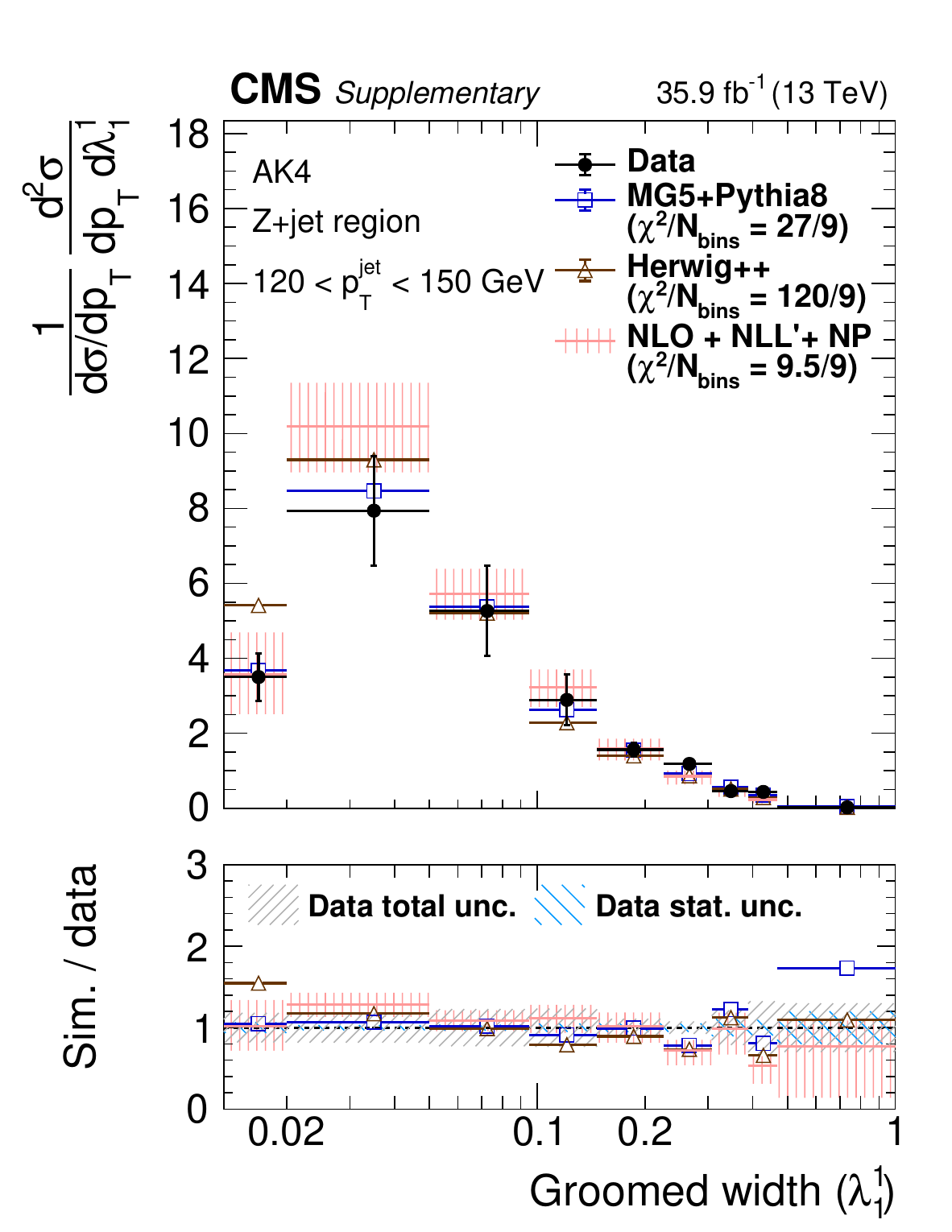}}
\caption{Groomed generalized angularities in $Z+jet$ events. From left to right: Les Houches angularity ($\lambda_{0.5}^1$), thrust ($\lambda_{2}^1$) and groomed width ($\lambda_{1}^1$).}
\label{fig2}
\end{figure}

The groomed and ungroomed generalized angularities in dijet and $Z+jet$ events were also measured by the CMS Collaboration~\cite{cms1}. The idea is to increase the $\beta$ value for fixed $\kappa$ to increase the weight of angular effects. The more weight is given to angular scales, the better agreement of theory with data is found for ungroomed measurements~\cite{cms1}. The groomed measurements are shown in Fig.~\ref{fig2}. Soft-drop grooming is used  to remove soft, wide angle radiation. Some
tension between the measurements and theory is found at small $\beta=$0.5, which might be related to hard collinear splittings description.

The CMS Collaboration also measured the dijet over $Z+$jet ratio for the different angular observables benefitting from the fact that many experimental uncertainties partially cancel in the ratio.
Leading order predictions and parton showers overestimate the
gluon to quark-enriched ratio. This ratio is better modelled with the ``old"
PYTHIA8 and HERWIG7 CMS
tunes~\cite{cms1}.
Angular measurements are fundamental to tune further MC and to understand better gluon radiation from QCD.

\section{Measuring the Lund jet plane using CMS data}

\begin{figure}
\centerline{%
\includegraphics[width=11.cm]{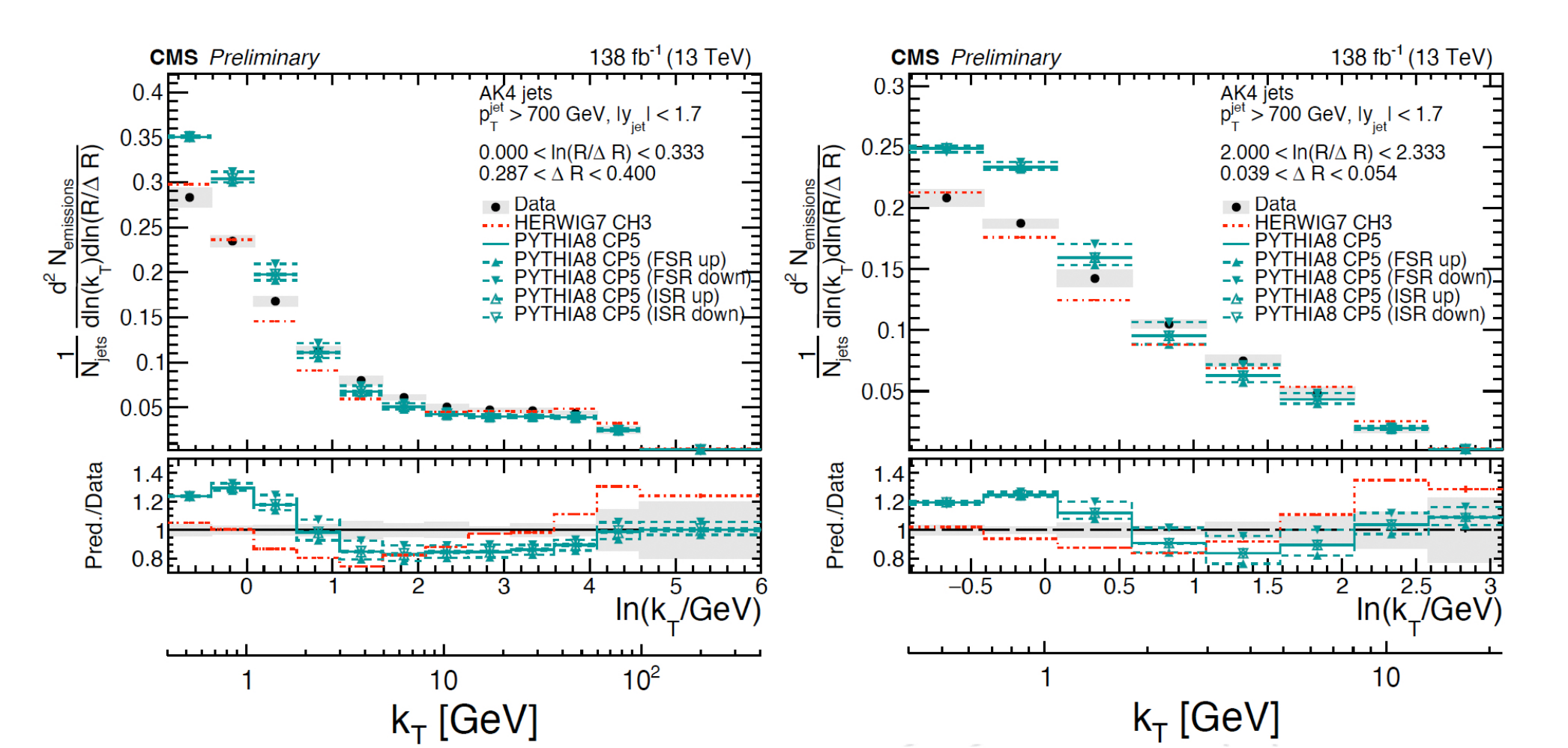}}
\caption{Lund jet plane density: $\log k_T$ dependence for two regions in $\Delta R$ and $\ln (R/\Delta R)$.}
\label{fig3}
\end{figure}

The idea of measuring the Lund jet plane in CMS is to visualize the phase-space of 1$\rightarrow$2 QCD splittings.
We define the
relative transverse momentum of the emission $k_T$ and the splitting angle $\Delta R= \sqrt{(y_{soft}-y_{hard})^2+(\phi_{soft}-\phi_{hard})^2}$.
Theoretically, Lund jet planes are used for parton shower calculations and jet substructure
techniques developments and experimentally, it is  possible to construct a
proxy for Lund diagrams using the iterative jet declustering procedure~\cite{gregory}.
In CMS, the constituents of the anti-$k_T$ jets are reclustered using
the Cambridge/ Aachen  (CA) algorithm~\cite{ca}.
The CA algorithm sequentially combines the pairs of protojets with
strict angular ordering, and the CA jet is then declustered iteratively (from large to
small angle emissions).
The transverse momentum $k_T$ and splitting angle $\Delta R$ of
soft subjet emission relative to the hard subjet
(core) are measured at each step
$k_T = p_T \Delta R$
where $p_T$ is the subjet transverse momentum. The procedure is iterated until the core is a single particle.

The measurements of the primary Lund jet plane were performed for jet radius $R=0.4$ and for the first time for $R=0.8$ and corrected to particle level~\cite{cms2}.
The primary Lund jet plane density projected onto the $\log k_T$ axis is shown in Fig.~\ref{fig3} for 0.4 jets (the results for 0.8 jets can be found in Ref.~\cite{cms2}).  The results for large (respectively small) splitting angles are shown on the  left (respectively right) plot.
PYTHIA8 CP5~\cite{pythia8} overestimates the number of emissions by 15-20\%, Data favor lower final state radiation  in the parton shower region. HERWIG 7 CH3~\cite{herwig7} leads to a better agreement with data.

\begin{figure}
\centerline{%
\includegraphics[width=11.cm]{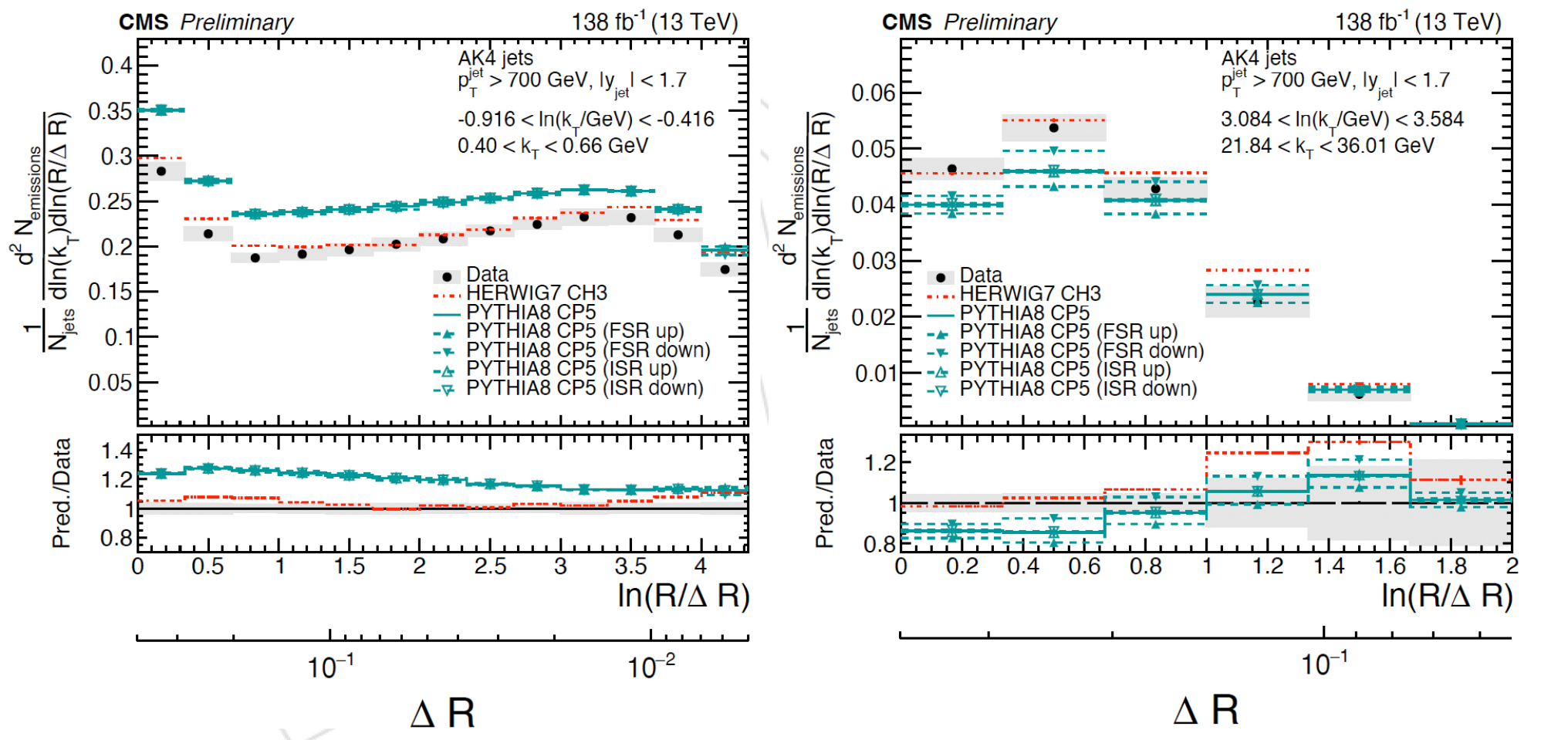}}
\caption{Lund jet plane density: $\log R/\Delta R$ dependence for two regions in $\Delta R$ and $\ln (R/\Delta R)$.}
\label{fig4}
\end{figure}

The primary Lund jet plane projected onto the $\log R/\Delta R$ axis is shown in Fig.~\ref{fig4} for $R=0.4$.
The region corresponding to soft (respectively hard) splittings is shown on the left (respectively right) plot. 
Low $k_T$ splitting populates the whole angle radiation region while the large $k_T$ only the wide one. 
PYTHIA8 CP5 overshoots data by 25-35\% at low $k_T$ and leads to a  better description for hard emissions.

\begin{figure}
\centerline{%
\includegraphics[width=10.cm]{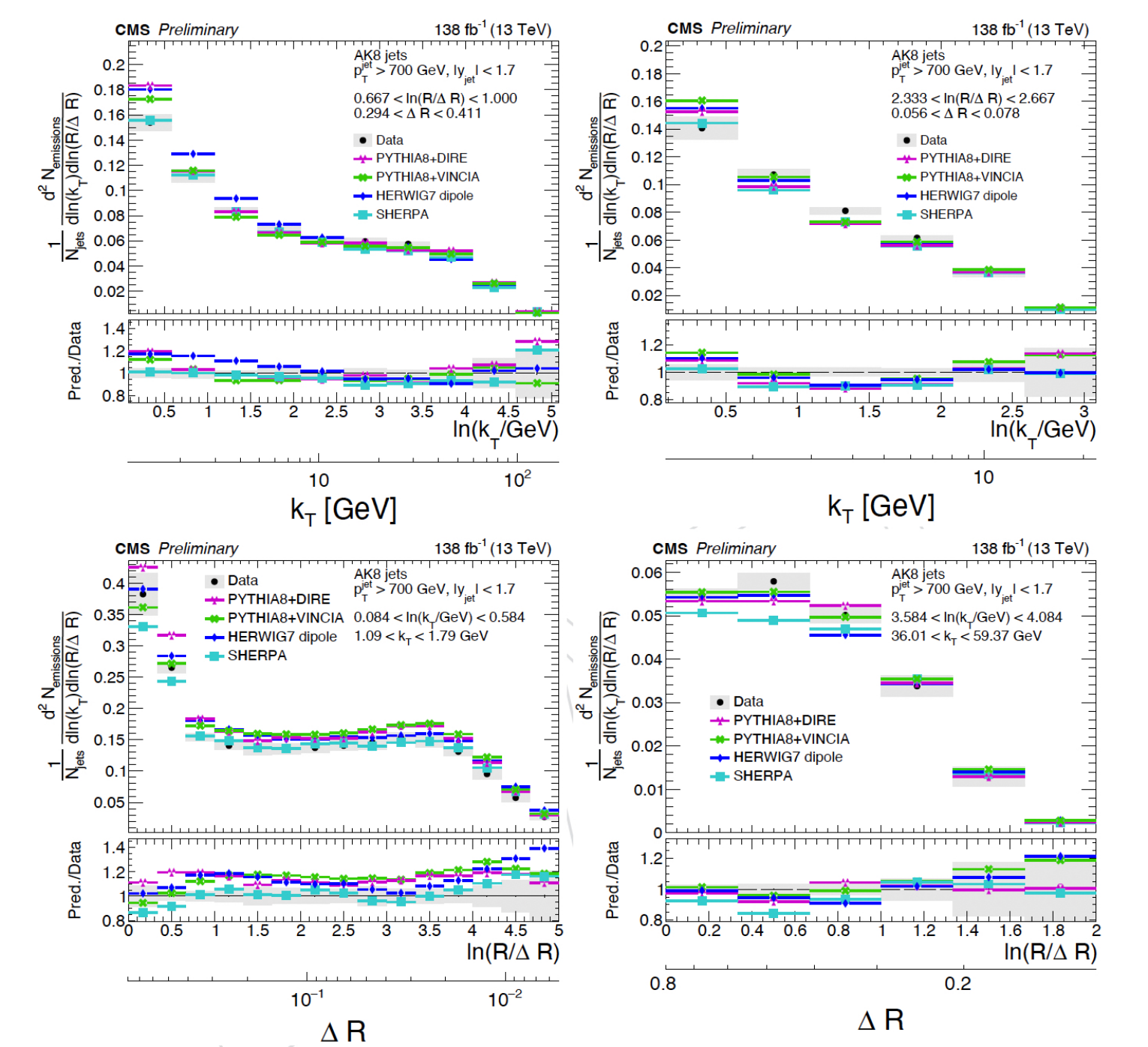}}
\caption{Lund jet plane density: comparison with the expectations from PYTHIA, HERWIG and SHERPA as a function of jet $k_T$ and $\Delta R$ for different regions  in $\Delta R$ and $\ln (R/\Delta R)$.}
\label{fig5}
\end{figure}

The comparison between the Lund jet plane measurement and model expectations is shown in Fig.~\ref{fig5} for different region in $\ln (R/\Delta R)$ and $\Delta R$. 
PYTHIA8 with VINCIA~\cite{vincia} or DIRE~\cite{dire} models are  in agreement with data within a few \% except at high $k_T$.
SHERPA~\cite{sherpa} and HERWIG7~\cite{herwig7} with dipole showers describe the data within 5-10\% including at high $k_T$.
The comparison between data and HERWIG7 allows to
choose the best recoil  scheme in angular ordered parton showers in a region where quark and gluon fragmentations play an important role.
The ultimate goal will be to achieve NLL accuracy in the next generation of parton showers.

The measurement of the jet-averaged density of emissions also scales with $\alpha_S$ 
in the soft and collinear limit of perturbative QCD, which shows that the measurement of jet substructure and 
of the Lund jet plane is directly sensitive to the value of $\alpha_S$~\cite{cms2}. 
The density of emissions has a simple dependence at leading log
$ 2/\pi  C  alpha_S (k_T)$, where C is the color factor.
One can use the Lund plane density to calibrate $\alpha_S$ from the parton shower (as a MC tuning parameter) since this is what drives part of the differences between the MC generators.

To conclude, we described two measurements of jet substructure that are sensitive to basic building blocks of QCD.
The measurement of the angular distributions in  Z$+$jet and dijet events are a valuable input for a better understanding of quark-jet and gluon-jet substructure. In addition, the measurement of the Lund jet plane is crucial to visualize the phase space of QCD splittings and it will improve our understanding of QCD and the description of data by MC, with the goal of achieving NLL accuracy in the next generation of parton showers.

\end{document}